\documentclass{elsart3p}
\usepackage{times,mathptmx}
\usepackage[numbers,sort&compress]{natbib}
\usepackage{amssymb}
\usepackage[final]{graphicx}
\usepackage{units}
\usepackage{amsmath}

\def\lsi{\raise0.3ex\hbox{$<$\kern-0.75em\raise-1.1ex\hbox{$\sim$}}}
\def\gsi{\raise0.3ex\hbox{$>$\kern-0.75em\raise-1.1ex\hbox{$\sim$}}}

\def\be{\begin{equation}}
\def\ee{\end{equation}}
\def\ba{\begin{eqnarray}}
\def\ea{\end{eqnarray}}

\begin{document}


\begin{frontmatter}
\title{Asymptotic safety of gravity and the Higgs boson mass}

\author{Mikhail Shaposhnikov}
\address{
  Institut de Th\'eorie des Ph\'enom\`enes Physiques,
  \'Ecole Polytechnique F\'ed\'erale de Lausanne,
  CH-1015 Lausanne, Switzerland}
\author{Christof Wetterich}
\address{ Institut f\"ur Theoretische Physik, Universit\"at Heidelberg,
Philosophenweg 16, D-69120 Heidelberg, Germany}

\today

\begin{abstract}
There are indications that gravity is asymptotically safe. The
Standard Model (SM) plus gravity could be valid up to arbitrarily high
energies. Supposing that this is indeed the case and assuming that
there are no intermediate energy scales between the Fermi and Planck
scales we address the question of whether the mass of the Higgs boson
$m_H$ can be predicted. For a positive gravity induced anomalous
dimension $A_\lambda>0$ the running of the quartic scalar self
interaction $\lambda$ at scales beyond the Planck mass is determined
by a fixed point at zero.  This results in $m_H=m_{\rm min}=126$ GeV,
with only a few GeV uncertainty. This prediction is independent of the
details of the short distance running and holds for a wide class of
extensions of the SM as well. For $A_\lambda <0$ one finds $m_H$ in
the interval $m_{\rm min}< m_H < m_{\rm max}\simeq 174$ GeV, now
sensitive to $A_\lambda$ and other properties of the short distance
running. The case $A_\lambda>0$ is favored by explicit computations
existing in the literature. 
\end{abstract}


\end{frontmatter}


Though gravity is non-renormalizable by perturbative methods, it may
exist as a field theory non-perturbatively \cite{Weinberg1979},
exhibiting a non-trivial ultraviolet fixed point (FP) of the
functional  renormalization group flow
\cite{Wilson:1973jj,Polchinski:1983gv,Wetterich:1992yh}. In
\cite{Reuter:1996cp} such a fixed point was indeed found in the
so-called Einstein-Hilbert truncation. Many works (for a recent review
see \cite{Codello:2008vh}), based on the exact functional
renormalization group equation (FRGE) of \cite{Wetterich:1992yh} (for
a review see \cite{Berges:2000ew}), produced  further evidence in
favor of this conjecture. The non-perturbative FP of
\cite{Reuter:1996cp} stays in place when higher order operators are
added to Einstein-Hilbert action, when the form of the infrared cutoff
is changed, etc. A similar picture arises in lattice formulations of
quantum gravity \cite{Ambjorn:2004qm} (for a recent review see
\cite{Ambjorn:2009ts}). Yet another indication comes from perturbative
computations \cite{Niedermaier:2009}.

The ``flowing action'' or ``effective average action'' $\Gamma_k$
includes all quantum fluctuations with momenta larger than an infrared
cutoff scale. For $k\to\infty$ no fluctuations are included and
$\Gamma_{k\to\infty}$ coincides with the classical or microscopic
action, while for $k\to 0$ the flowing action includes all quantum 
fluctuations and becomes the generating functional of the one-particle
irreducible Green's functions. The scale dependence of $\Gamma_k$
obeys an exact functional renormalization group equation
\cite{Wetterich:1992yh}. It is of a simple one loop type, but
nevertheless can be solved only approximately by suitable
non-perturbative truncations of its most general functional form.

From the studies of the functional renormalization group for
$\Gamma_k$ one infers a characteristic scale  dependence of the
gravitational constant or Planck mass, 
\be
\label{AA}
M_P^2(k) = M_P^2 +2 \xi_0 k^2~,
\ee
where $M_P = (8\pi G_N)^{-1/2}= 2.4 \times 10^{18}$ GeV is the low
energy Planck mass, and $\xi_0$ is a pure number, the exact value of
which is not essential for our considerations. From investigations of
simple truncations of pure gravity one finds $\xi_0\approx 0.024$ from
a numerical solution of FRGE \cite{Reuter:1996cp,
Percacci:2003jz,Narain:2009fy}. For scattering with large momentum
transfer $q$ the effective infrared cutoff $k^2$ is replaced by $q^2$.
Thus for $q^2\gg M^2_p$ the effective gravitational constant $G_N
(q^2)$ scales as $\frac{1}{16\pi\xi_0 q^2}$, ensuring the regular
behavior of high energy scattering amplitudes. 

We can distinguish two qualitatively different regimes, separated by a
transition scale
\be
\label{trans}
k_{tr}=\frac{M_P}{\sqrt{2\xi_0}}\approx 10^{19}~{\rm GeV}.
\ee
For the ``high energy regime'' $k\gtrsim k_{tr}$ we observe scaling
behavior  $M^2_P(k)/k^2\approx 2\xi_0$, characteristic for an
ultraviolet fixed point. In contrast, for the ``low energy regime''
the effects of graviton loops are effectively switched off and
$M^2_P(k)$ becomes a scale independent constant. Eq. \eqref{AA}
describes the typical behavior for a ``relevant parameter''
characterizing a deviation from an exact fixed point. We observe that
the high energy regime is essentially determined by canonical
dimensional scaling in absence of mass scales. We expect this form to
hold for a wide class of theories with an ultraviolet fixed point,
where the high energy regime may involve additional fields or even
higher dimensions. The numerical value of $\xi_0$ will then depend on
the precise model. (In the presence of an anomalous dimension for the
graviton eq. \eqref{AA} holds for an appropriate renormalized coupling
$M^2_P(k)$.) In the ``low energy regime'' $k\lesssim k_{tr}$ the
running of the gravitational couplings is essentially stopped. 

For $k\gtrsim k_{tr}$ the running of the dimensionless couplings of
the Standard Model is modified by gravitational contributions. We may
denote these couplings by $x_j$ for the gauge couplings $g_1,~g_2,~
g_3$ of $U(1),~SU(2),~SU(3),~h$ for the top Yukawa coupling and
$\lambda$ for the self interaction of the Higgs scalar. The
gravitational contribution to the beta-functions $\beta^{grav}_j$
takes typically the form\footnote{We would like to stress that the
definition of the running couplings here is based on the
gauge-invariant high energy physical scattering amplitudes
\cite{Weinberg1979}, rather than on the minimal subtraction (MS)
scheme of the dimensional regularization. In the MS scheme
perturbative Einstein gravity does not contribute to the $\beta$
functions of the Standard Model couplings \cite{Ebert:2007gf}.}
\be
\label{1A}
\beta^{grav}_j=\frac{a_j}{8\pi}\frac{k^2}{M^2_p(k)}x_j~.
\ee
For the high energy regime this amounts to effective anomalous
dimensions
\be
\label{1B}
A_j=\frac{a_j}{16\pi\xi_0}~.
\ee
For small $x_j$ eq. \eqref{1A} describes the leading contribution,
such that 
\be
\label{1C}
x_j(k)\sim k^{A_j}~.
\ee

The general form \eqref{1A} is again dictated by simple scaling
arguments. Explicit computations confirm these expectations
\cite{Percacci:2003jz,Narain:2009fy,Daum:2009dn,Robinson:2005fj,
Pietrykowski:2006xy,Toms:2007sk,Toms:2008dq,Zanusso:2009bs}. In
general, the constants $a_j$ will depend on the precise model which
describes the high energy regime. We emphasize that for $a_j<0$ the
running of $x_j$ is asymptotically free, at least for small enough
values of the coupling. For the low energy regime $k^2\lesssim
k^2_{tr}$ the gravitational contributions become negligible. 

Within this setting a very economical description of all interactions
in Nature may be possible. One can assume that there is no new physics
associated with any intermediate energy scale (such as Grand Unified
scale or low energy supersymmetry) between the weak scale and
$k_{tr}$. All confirmed observational signals in favor of physics
beyond the Standard Model as neutrino masses and oscillations, dark
matter and dark energy, baryon asymmetry of the Universe and
inflation  can be associated with new physics {\em below} the
electroweak scale, for reviews see
\cite{Shaposhnikov:2007nj,Boyarsky:2009ix} and references therein. The
minimal model -- $\nu$MSM, contains, in addition to the SM particles,
3 relatively light singlet Majorana fermions and the dilaton. These
fermions could be responsible for neutrino masses, dark matter and
baryon asymmetry of the Universe. The dilaton may lead to dynamical
dark energy \cite{Wetterich:1987fm,Wetterich:1987fk} and realizes
spontaneously broken scale invariance which either emerges from the
cosmological approach to a fixed point \cite{Wetterich:1987fm,16A} or
is an exact quantum symmetry
\cite{Shaposhnikov:2008xb,Shaposhnikov:2008xi}. Inflation can take
place either due to the SM Higgs \cite{Bezrukov:2007ep} or due to the
asymptotically safe character of gravity \cite{Weinberg:2009wa}. Yet
another part of new physics, related, for example, to the strong CP
problem or to the flavor problem, may be associated with the Planck
energy.  In this Letter we show that this scenario leads to a
prediction of the Higgs mass, which can be tested at the LHC.

A convenient language for understanding the origin of this prediction
is the concept of infrared intervals \cite{20A}. Consider first the
low energy regime where graviton loops can be neglected and the $x_j$
follow the perturbative renormalization group equations of the SM,
$k\partial x_j/\partial_k=\beta^{\rm SM}_j$, with one loop expressions
\begin{align}
\nonumber
\beta_1^{\rm SM}&=\frac{41}{96\pi^2} g_1^3\;,\;\;\;
\beta_2^{\rm SM}=-\frac{19}{96\pi^2}g_2^3 \;,\;\;\;
\end{align}
\begin{align}
\label{betafunctions}
\beta_3^{\rm SM}&=-\frac{7}{16\pi^2}g_3^3\;,\\
\beta_h^{\rm SM}&=\frac{1}{16\pi^2}\left[\frac{9}{2}h^3-8g_3^2h-
\frac{9}{4}g_2^2h-\frac{17}{12}g_1^2h\right]\;,\\
\notag
\beta_\lambda^{\rm SM}&=\frac{1}{16\pi^2}
\left[24\lambda^2+12\lambda
h^2-9\lambda(g_2^2+\frac{1}{3}g_1^2)\right.\\
&\left.-6h^4+\frac{9}{8}g_2^4+\frac{3}{8}g_1^4
+\frac{3}{4}g_2^2g_1^2\right]~.
\label{betalambda}
\end{align}

The ratio $\lambda/h^2$ has a partial infrared fixed point
\cite{20A,20B} (up to small modifications due to the gauge couplings).
Since the running within the low energy regime extends only over a
finite range between $k_{tr}$ and the Fermi scale $k_F$, this fixed
point needs not to be approached arbitrarily close. Instead, arbitrary
couplings at the scale $k_{tr}$ in the allowed range $0\leq
h^2(k_{tr})<\infty,~0\leq \lambda(k_{tr})\leq \infty$ are mapped by the
RG-flow to an infrared interval of allowed couplings at the Fermi
scale $k_F$. For known top quark mass or fixed $h(k_F)$ the infrared
interval for $\lambda(k_F)/h^2(k_F)$, centered around the partial
fixed point, determines the allowed values of the mass of the Higgs
doublet. The upper limit $\lambda_{\rm max}(k_F)$ corresponds to the
``triviality bound''. Numerically, it coincides essentially with the
requirement that for $k<k_{tr}$ the SM-coupling should remain within the
perturbative range \cite{Maiani:1977cg,Cabibbo:1979ay}, but its
validity extends beyond perturbation theory \cite{20C}. The lower
limit $\lambda_{\rm min}(k_F)$ arises from the observation that even
for $\lambda(k_{tr})=0$ a nonzero $\lambda(k_F)$ is generated due to
the term $\sim h^4$ in $\beta_\lambda$. An extended range of large
$\lambda(k_{tr})$ is mapped to $\lambda(k_F)$ close to $\lambda_{\rm
max}(k_F)$, while a large range of small $\lambda(k_{tr})$ is mapped
to values of $\lambda(k_F)$ close to the lower bound $\lambda_{\rm
min}(k_F)$ \cite{20A}. This observation will be crucial for our
prediction.

We next discuss the running in the high energy regime. The allowed
values of $x_j(k_{tr})$ correspond now to the infrared interval for
the first stage of the running. Since we want this running to hold for
arbitrarily large $k$, the infrared intervals are completely
determined by the possible fixed points. If some coupling or ratio of
couplings has only one infrared stable fixed point, the value at
$k_{tr}$ must be given by the fixed point value and becomes
predictable. In case of an ultraviolet fixed point only, the value of
the coupling at the transition scale $k_{tr}$ remains undetermined,
since arbitrary values of $x_j(k_{tr})$ run to the ultraviolet fixed
point for $k\to\infty$. Finally, we consider the case where a coupling
or combination of couplings $x$ has an infrared stable fixed point at
$x_{IR}$ and a second ultraviolet stable fixed point at $x_{UV}$. For
$x_{IR}<x_{UV}$ the infrared interval for $x$ is given by
$x(k_{tr})\geq x_{IR}$, while for $x_{IR}>x_{UV}$ one finds
$x(k_{tr})\leq x_{IR}$. 

The most interesting situation for a prediction of the mass of the
Higgs scalar arises if $h^2$ has an ultraviolet fixed point
$h^2_{UV}=0$, while $\lambda$ has an infrared fixed point
$\lambda_{IR}=0$ in the limit where $h$ and $g_i$ vanish for $k\gg
k_{tr}$. This setting is realized for $a_h<0,~a_\lambda>0$. In this
case $\lambda(k_{tr})$ is predicted very close to zero, such that
$\lambda(k_F)$ will be very close to the lower bound $\lambda_{\rm
min}(k_F)$. This results in a Higgs-scalar mass $m_H\approx 126$ GeV,
see below. We emphasize that only inequalities for a $a_h$ and
$a_\lambda$ are needed for this prediction, while the precise value of
these constants does not matter. It is also essential that the sign of
the gravity contribution to the running of all gauge couplings is
negative, $a_i<0$. 

To substantiate the general discussion given above, consider the pure
SM coupled to gravity, with running couplings given by
\be
\label{2}
k\frac{dx_j}{dk}=\beta^{\rm SM}_j+\beta^{grav}_j~.
\ee
As for the gauge couplings, we will fix their values at small energies
to the experimental ones, but will leave  $\lambda$ and $h$
undetermined for the time being. 

First, let us look at the gauge sector. Assume for simplicity that
$a_1=a_2=a_3=a_g$, which is true for one-loop computations, performed
up to now, due to the universality of the gravitational interactions.
For $a_g<0$ all gauge couplings are asymptotically free. Indeed, the
computations of \cite{Robinson:2005fj,Daum:2009dn} yield a {\em
negative} sign for $a_g$, with $|a_g|\sim 1$. In this case the gauge
coupling constants $g_2$ and $g_3$ cannot be predicted. The gauge
coupling $g_1$ has two fixed points
\be
\label{6A}
g^2_{1,UV}=0~,~g^2_{1,IR}=\frac{6\pi|a_g|}{41\xi_0}~,
\ee
such that $g^2_1(k_{tr})\leq g^2_{1,IR}$. For a realistic gauge
coupling one needs $g_1(k_{tr})\approx 0.5$, and this requires
$a_g<a^{\rm crit}_g\approx -0.013$. Then the Landau pole problem for
the $U(1)$-coupling is solved due to the presence of the fixed point.
We will assume in the following $a_g<a^{\rm crit}_g$ and take for
definiteness the value $|a_g| \sim 1$. For large enough $k\gg k_{tr}$ the
gauge couplings can be neglected for the running of $h$ and
$\lambda$. 

Consider now the top Yukawa coupling $h$. For a positive anomalous
dimension $a_h>0$ one finds only an $IR$-stable fixed point at
$h^2_{IR}=0$. This would predict $h^2(k_{tr})=0$, and therefore a
vanishing top quark mass. Clearly, this case is rejected by
experiment. For the interesting case $a_h<0$ there are two fixed
points
\be
\label{6B}
h^2_{UV}=0~,~h^2_{IR}=\frac{2\pi|a_h|}{9\xi_0}~,
\ee
implying $h(k_{tr})\leq h_{\rm max}(k_{tr})\approx h_{IR}$. (There are
small numerical modifications of the second relation due to the
presence of the gauge couplings.) We may compute numerically the value
of $h(k_{tr})$ which corresponds to the (central) experimental value
of the top quark mass, $m_t=171.3$ GeV \cite{Amsler:2008zzb}. Since
this has to be smaller than $h_{\rm max}(k_{tr})$ a realistic setting
requires $a_h<a^{\rm crit}_h\approx -0.005$. An interesting scenario
would arise if $h(k)$ gets close to the fixed point value $h_{IR}$ for
$k\gg k_{tr}$. In this case the top quark mass becomes predictable,
and a realistic value requires  $a_h=a^{\rm crit}_h,~h_{IR}\approx
0.38$. 

At present, the value of $a_h$ is not known reliably. For example, in 
\cite{Shapiro:1989dq} it was shown that gravity contributions make
the  Yukawa coupling asymptotically free in quantum $R^2$ gravity with
matter. Ref. \cite{Zanusso:2009bs} studied  the gravitational running
of Yukawa couplings in the FRGE approach for the Einstein-Hilbert type
of truncation and found different signs for $a_h$ in different gauges.
In this work the wave function renormalization for the fermions and
scalars was not included and the sensitivity to the truncation type
was not investigated. In what follows we will simply assume that $a_h
< a_h^{\rm crit}$ which is the only realistic case for observations.

Let us turn now to the behavior of the scalar self-coupling $\lambda$.
The gravitational corrections can only promote the SM to the rank of
fundamental theory if the running of $\lambda$ does not lead to any
pathologies up to the Planck scale. In other words, the Landau pole
must be absent for $k \lsi k_{tr}$
\cite{Maiani:1977cg,Cabibbo:1979ay,Lindner:1985uk}, and $\lambda$ must
be positive for all momenta up to $k_{tr}$
\cite{Krasnikov:1978pu,Hung:1979dn,Politzer:1978ic}. There is a large
parameter space available on the plane $m_H,m_t$, where both
conditions are satisfied. Close to the experimental value of the top
mass, it is described by the infrared interval for $\lambda(k_F)$,
corresponding to $m_{\rm min}< m_H < m_{\rm max}$. Here
\ba
\nonumber
m_{\rm min}= [126.3 + \frac{m_t - 171.2}{2.1}\times 4.1\\
-\frac{\alpha_s-0.1176}{0.002}\times 1.5]~{\rm GeV}~,
\label{mmm}
\ea
and 
\ba
\nonumber
 m_{\rm max} = [173.5 +\frac{m_t - 171.2}{2.1}\times 0.6\\
 -\frac{\alpha_s-0.118}{0.002}\times 0.1]~ {\rm GeV}~,
 \label{11}
\ea
where  $\alpha_s$ is the strong coupling at the $Z$-mass, with
theoretical uncertainty in $m_{\rm min}$ equal to $\pm 2.2$ GeV. These
numbers are taken from the recent two-loop analysis of 
\cite{Bezrukov:2009db}  (see also  \cite{Espinosa:2007qp,Ellis:2009tp}
and earlier computations in
\cite{Altarelli:1994rb,Casas:1994qy,Casas:1996aq,Hambye:1996wb}). The
value of $m_{\rm max}$ corresponds to the (somewhat arbitrary)
criterion $\lambda(k_{tr}) < 6$, but changes only very little for
arbitrarily large $\lambda(k_{tr})$. The admitted region contains also
very small top and Higgs masses, excluded experimentally.

As we have already said, a specific prediction of the Higgs boson mass 
can be given if $a_\lambda$ is positive. In fact, the evidence that
this is indeed the case for the SM coupled to gravity comes from
computations of \cite{Percacci:2003jz,Narain:2009fy}, giving
\be
a_\lambda \approx 3.1~,~A_\lambda\simeq 2.6~.
\ee
A contribution with the same sign and similar magnitude was found
previously in  \cite{Griguolo:1995db}.

Let us elucidate the structure of the solution to the RG equation for
$\lambda$ in this case. For $a_h<a^{\rm crit}_h$, $a_g<a^{\rm
crit}_g$, asymptotic freedom of the gauge and Yukawa couplings implies
that they can be neglected for $k\gg k_{tr}$. (We assume here for
simplicity negative values of $a_h,~a_g$ of the order one, such that
this regime is reached for scales only moderately above $k_{tr}$.) The
remaining terms in $\beta_\lambda$ drive then $\lambda$ quickly to an
approximate infrared fixed point $\lambda_{IR}=0$. This is the only
fixed point such that $\lambda(k_{tr})$ becomes predictable. The
actual value $\lambda(k_{tr})$ differs from zero only due to the
presence of nonzero gauge and Yukawa couplings. For large enough $h$
the term $\sim -h^4$ in $\beta_\lambda$ dominates over the terms $\sim
g^4$, such that $\lambda$ is driven to a small positive value. For
realistic values of $h(k_{tr})$ and $g_i(k_{tr})$ the effects driving
$\lambda$ away from zero are small, however, and they act only in a
small region $k\gtrsim k_{tr}$ before $h(k)$ and $g_i(k)$ drop to very
small value for larger $k$. As a result, $\lambda(k_{tr})$ remains
very small, such that $\lambda(k_F)$ is predicted very close to the
lower bound of the infrared interval. This yields a robust prediction
$m_H=m_{\rm min}$, independently of the precise values of the
constants $a_j$!

Indeed, for this scenario the trajectory of $\lambda(k)$ is given by
the special solution 
\ba
\label{lfp}
\lambda(k)=\;\;\;\;\;\;\;\;\;\;\;\;\;\;\;\;\;\;\;\;\;\;\;\;\;\;\;\;\;\;\;\;
\;\;\;\;\;\;\;\;\;\;\;\;\;\;\;\;\;\;\;\;\;\;\;\;\;\\
-\int_k^\infty \frac{dk'}{k'}
\left(\frac{1+2\xi_0 k^2/M_P^2}{1+2\xi_0
k'^2/M_P^2}\right)^{\frac{a_\lambda}{32\pi\xi_0}}
\times\beta_\lambda^{\rm SM}(x_j(k'))~.
\nonumber
\ea

For $a_\lambda>0$ only a small range of $k'$ in the vicinity of $k$
contributes to the integral in eq. \eqref{lfp}. We infer
\be
\label{16A}
\lambda(k_{tr})=-C\beta^{\rm SM}_\lambda\big (h(k_{tr}),g_i(k_{tr})\big),
\ee
where $C$ is positive and of order one. Since $\lambda(k_{tr})$ is
small we can omit the terms involving $\lambda$ in
$\beta^{\rm SM}_\lambda$, such that $\lambda(k_{tr})$ is fixed in terms of
$h(k_{tr})$ and $g_i(k_{tr})$. 

While the effects of the terms $\sim h^4,~g^4$ in $\beta_\lambda$ are
numerically small, they impose nevertheless an important lower bound
for the allowed values of the top quark mass. For too small values of
$h$ the positive contributions $\sim g^4$ will dominate over the
negative values $\sim -h^4$ in $\beta_\lambda$. Since the terms $\sim
\lambda a_\lambda$ and $\sim \lambda^2$ drive $\lambda(k)$ quickly
towards zero as $k$ is lowered, a remaining positive $\beta_\lambda$
for $\lambda=0$ would induce $\lambda(k)$ to run to negative values.
Such a behavior can be associated with radiatively induced spontaneous
symmetry breaking of the Coleman-Weinberg \cite{42A} type, at a scale
close to $k_{tr}$ or above. A realistic scenario of electroweak
symmetry breaking, supplemented by cosmological considerations such as
Higgs-inflation  \cite{Bezrukov:2007ep}  or asymptotically safe
inflation \cite{Weinberg:2009wa}, has to exclude this case, therefore
requiring that $\lambda(k)$ remains positive for all values of $k$. 
From numerical solutions of the RG-equations we infer a lower bound
for the top quark mass
\be
\label{BA}
m_t\geq m_t^{\rm min}~,
\ee
where $m_t^{\rm min}\simeq 170$ GeV,  slightly depending on the values of
anomalous dimensions (for example, $a_g=a_h=-1,~a_\lambda=3$ gives
$m_t^{\rm min} \simeq 173.4$ GeV, whereas
$a_g=-1,~a_h=-0.5,~a_\lambda=3$ leads to $m_t^{\rm min} \simeq 169.3$
GeV). It is interesting that the experimental value $m_t=171.3$ GeV is
very close to this lower limit. This implies that both $\lambda$ and
$\beta_\lambda$ are close to zero at the transition scale $k_{tr}$,
\be
\label{BB}
\lambda(k_{tr})\approx 0~,~\beta_\lambda(k_{tr})\approx 0~.
\ee
This suggests that the fundamental theory may be characterized by a
fixed point at $\lambda=0$ also for nonzero $h$ and $g$, thereby
predicting $m_t$ to be close to $170$ GeV. Furthermore, the
requirement that the Yukawa contribution to $\beta_\lambda$ continues
to dominate over the gauge boson contributions for very large $k$
imposes a constraint
\be
a_g \leq a_h \leq a_h^{\rm crit}~.
\label{yukawaconstr}
\ee
If this condition holds, we find a range of $k$ (larger than $k_{tr}$)
for which the running of $\lambda$ can be approximated by the
simplified equation
\be
\label{14A}
\beta_\lambda=\frac{a_\lambda}{16\pi\xi_0}\lambda+\frac{1}{16\pi^2}
(24\lambda^2+12\lambda h^2-6h^4)~.
\ee
For a fixed point behavior of the Yukawa coupling $a_h=a^{\rm
crit}_h~,~h=h_*=h_{IR}$ \eqref{6B} this yields a fixed point for
$\lambda$ obeying 
\be
\label{16}
24{\lambda}^2_*+12\lambda h^2_*-6{h^4_*}+
\frac{\pi a_\lambda\lambda_*}{\xi_0}=0~.
\ee

These findings can be verified by an explicit numerical solution of
the RG-equations \eqref{2}. For better accuracy, in the numerical
computations we used for $\beta^{\rm SM}_j$ the two-loop RG equations and
one-loop pole matching of the physical parameters, see
\cite{Luo:2002ey,Espinosa:2007qp} and also \cite{Bezrukov:2009db}. We
run the normalization flow towards the ultraviolet by increasing $k$,
starting at the Fermi scale $k_F$ with the gauge couplings as inferred
from experiment and with a given fixed $m_t$. For $a_g<0~,~a_g\leq
a_h\leq a^{\rm crit}_h~,~a_\lambda>0$ we find that indeed only a
single value of $\lambda(k_F)$ can be extrapolated to arbitrarily
large $k$, while larger values diverge and smaller values turn
negative. This corresponds to the prediction that the infrared
interval consists only of one point, corresponding to the approximate
fixed point $\lambda_{IR}=0$ for sufficiently large $k$ where $h^2$
and $g^2$ can be neglected, or to the value $\lambda_*$ in eq.
\eqref{16} if $a_h=a^{\rm crit}_h$. This solution exists only provided
the bound for $m_t$ \eqref{BA} is obeyed. For example, for
$a_g=-1,~a_h=-0.5$, and $a_\lambda=3$ the admitted RG trajectories
exist for a large variety of top  masses: $m_t=171.3$ GeV leads to
$m_H \simeq 126.5$ GeV, whereas $m_t=230$ GeV requires  $m_H\simeq
233$ GeV. 

Let us now choose the experimental value for the top quark mass and
determine the Higgs boson mass which corresponds to the allowed value
of $\lambda(k_F)$. As expected, the prediction is quite insensitive to
the specific values of $a_g$, $a_h$ and $a_\lambda$ and reads
\be
m_H = m_{\rm min}~.
\label{predict}
\ee
The value of $m_H$ can only increase if the top Yukawa coupling is
close to its  non-Gaussian fixed point, $h_{IR}$, realized for
$a_h=a^{\rm crit}_h$, which leads to the existence of the non-trivial
fixed point in $\lambda$ \eqref{16}. Taking, as an example, $a_g=-1$,
$a_h^{\rm crit}\simeq -0.005$, one gets from (\ref{16})
$\lambda_*=0.043$. This shifts up the prediction of the Higgs mass by
not more than $8$ GeV. Taking smaller $|a_g|$ decreases this
shift\footnote{The values of the Higgs mass we found are consistent
with a possibility of inflation due to the SM Higgs boson
\cite{Bezrukov:2007ep}. The Higgs-inflation requires the consistency
of the SM up to the lower, than $M_P$ energy scale $k \sim
\frac{M_P}{\xi}$, where $\xi = 700-10^5$ is the value of the
non-minimal coupling of the Higgs field to the curvature Ricci scalar 
\cite{Bezrukov:2008ej,Bezrukov:2009db} (see also
\cite{DeSimone:2008ei,Barvinsky:2009fy}), the smaller $\xi$ correspond
to smaller Higgs masses.}.

The prediction (\ref{predict}) can be tested at the LHC\footnote{The
fact that the SM scalar self-coupling is equal to zero {\em together}
with its $\beta$-function at the Planck scale for the particular values
of the top-quark and Higgs masses was first (to the best of our
knowledge) noticed  in  \cite{Froggatt:1995rt}. These authors put
forward the hypothesis of a ``multiple point principle'', stating that
the effective potential for the Higgs field must have two minima, the
one corresponding to our vacuum, whereas another one must occur at the
Planck scale. Our reasoning is completely different. Though the sense
of the ``multiple point principle'' remains unclear to us, we would
like to note that the prediction of the Higgs mass from it coincides
with ours  (the specific numbers in \cite{Froggatt:1995rt} are
different, as they were based on one-loop computation).}. Given the
fact that the accuracy in the Higgs mass measurements at the LHC can
reach $200$ MeV, the reduction of theoretical uncertainty and of
experimental errors in the determination of the top quark mass and of
the strong coupling constant are highly desirable.  As was discussed
in \cite{Bezrukov:2009db}, the theoretical error can go down from
$2.2$ GeV to $0.4$ GeV if one upgrades the one-loop pole matching at
the electroweak scale and two loop running up to the Planck scale to
the  two-loop matching and 3-loop running. Note that 3-loop
beta-functions for the SM are not known by now, and that the two-loop
pole matching has never been carried out. Clearly,  computations of
signs and magnitudes of gravitational anomalous dimensions $A_i$ are
needed to remove yet another source of uncertainties.  

The prediction $m_H\approx m_{\rm min}$ does not only hold for the
hypothesis that the SM plus gravity describes all the physics relevant
for the running of couplings. It generalizes to many extensions of the
SM and gravity, including possibly even higher dimensional theories.
Of course, the precision of the prediction gets weaker if a much
larger class of models is considered. Nevertheless, only two crucial
ingredients are necessary for predicting $m_H\approx m_{\rm min}$: (i)
Above a transition scale $k_{tr}$ the running should drive the quartic
scalar coupling rapidly to an approximate fixed point at $\lambda=0$,
only perturbed by small contributions to $\beta_\lambda$ from Yukawa
and gauge couplings. This is generically the case for a large enough
anomalous dimension $A_\lambda>0$. (ii) Around $k_{tr}$ there should
be a transition to the SM-running in the low energy regime. This
transition may actually involve a certain splitting of scales as
``threshold effects'', for example by extending the SM to a Grand
Unified theory at a scale near $k_{tr}$. It is sufficient that these
threshold effects do not lead to a rapid increase of $\lambda$ in the
threshold region. This will be the case if the $\lambda$-independent
contributions to $\beta_\lambda$ only involve perturbatively small
couplings in a threshold region extending over only a few orders of
magnitude.

A possible alternative to the above prediction appears if we have a
negative anomalous dimension $A_\lambda<0$. In this case the
approximate $IR$-fixed point $\lambda_{IR}$ for vanishing $h$ and $g$
is shifted away from the ``Gaussian fixed point'' $\lambda=0$. From
eqs. (\ref{1A},\ref{betalambda}) one finds for $h=g=0$
\be
\label{18A}
\lambda_{IR}=\frac{\pi|a_\lambda|}{24\xi_0}=
\frac{2\pi^2}{3}|A_\lambda|~.
\ee

In this case the $IR$-interval becomes $0\leq \lambda(k_{tr})\leq
\lambda_{IR}$. Again, nonzero $h,~g$ will slightly shift the infrared
interval. However, the value of $\lambda(k_{tr})$ depends now strongly
on the details of the running in the high energy regime, in particular
on the value of $\lambda_{IR}$ (or $A_\lambda$). Without a precise
knowledge of this running this alternative only predicts $m_H$ to be
in the $IR$-interval  $m_{\rm min}<m_H<m_{\rm max}$, with $m_{\rm
min}$ and $m_{\rm max}$ given by eqs. (\ref{mmm},\ref{11}).

Finally, we turn to the running of the mass parameter in the
Higgs-potential $\mu^2(k)$. So far, we have implicitly assumed that
the Fermi scale is fixed to its experimental value. For $g_i=0,~h=0$,
a vanishing Fermi scale corresponds to a second order phase transition
between a phase with spontaneous electroweak symmetry breaking and a
phase with unbroken (global) $SU(2)\times U(1)$ symmetry.  If we
choose $\mu^2(k_{tr})$ to correspond precisely to the phase transition
the Fermi scale will vanish. A second order phase transition
corresponds to an exact fixed point, for which an effective dilatation
symmetry of the low energy theory becomes realized \cite{20A,43A}.
(The scale transformations of these ``low energy dilatations'' keep
the Planck mass fixed and are therefore  different from a possible
fundamental dilatation symmetry.) At the phase transition, the running
of $\mu^2$ is given by a critical trajectory $\mu^2_*(k)$. 

Deviations from the critical trajectory,
$\delta\mu^2(k)=\mu^2(k)-\mu^2_*(k)$, behave as a relevant parameter
for the low energy running. The running of $\delta\mu^2$ is governed
by an anomalous dimension
\be
\label{ZA}
k\frac{\partial}{\partial k}\delta\mu^2=A_\mu\delta\mu^2~,
\ee
such that a small $\delta\mu^2$ remains small during the flow. The
small parameter $\delta\mu^2$ is natural in a technical sense - it is
associated to a small deviation from an (almost) exact symmetry, i.e.
the low energy dilatations \cite{43A}. An appropriate renormalization
group improved perturbation theory requires no fine tuning order by
order \cite{43A}, the anomalous dimension $A_\mu$ can be computed in a
perturbative series in the couplings.

An important question concerns the allowed ``$IR$-interval'' for
$\delta\mu^2(k_{tr})$. This will depend on the size and sign of
$A_\mu$ in the high energy regime. For a large positive $A_\mu$ one
infers that $\delta\mu^2(k_{tr})$ should be very close to zero. In
particular, for $A_\mu>2$ the dimensionless ratio $\delta\mu^2/k^2$ is
attracted to zero, corresponding to ``self organized criticality''
\cite{20B,43B}. This could help to understand the small ratio between
the Fermi and Planck scales. From presently published results
\cite{Percacci:2003jz,Narain:2009fy} for the scalar theory coupled
to gravity one infers $A_\mu=1.83$; what happens in the full SM is
unknown.

In conclusion, we discussed the possibility that the SM, supplemented
by the asymptotically safe gravity plays the role of a  fundamental,
rather than effective field theory. We found that this may be the case
if the gravity contributions to the running of the Yukawa and Higgs
coupling have appropriate signs. The mass of the Higgs scalar is
predicted  $m_H = m_{\rm min}\simeq 126$ GeV with a few GeV
uncertainty if all the couplings of the Standard Model, with the
exception of the Higgs self-interaction $\lambda$, are asymptotically
free, while $\lambda$ is strongly attracted to an approximate fixed
point $\lambda=0$ (in the limit of vanishing Yukawa and gauge
couplings) by the flow in the high energy regime. This can be achieved
by a positive gravity induced anomalous dimension for the running of
$\lambda$. A similar prediction remains valid for extensions of the SM
as grand unified theories, provided the split between the unification
and Planck-scales remains moderate and all relevant couplings are
perturbatively small in the transition region. Detecting the Higgs
scalar with mass around $126$ GeV at the LHC could give a strong hint
for the absence of new physics influencing the running of the SM
couplings between the Fermi and Planck/unification scales.

{\bf Acknowledgments.}
The work of M.S. was supported by the Swiss National Science
Foundation and by Alexander von Humboldt Foundation. M.S. thanks
Heidelberg University, were this work was done, for kind hospitality.
Helpful discussions with Fedor Bezrukov are appreciated.


\end{document}